\documentstyle[11pt]{article}

\newcommand{\partialslash}{\partial \!\!\! /}

\newcommand{\xslash}{x \!\!\! /}

\newcommand{\half}{\mbox{\small{$\frac{1}{2}$}}}

\begin{document}
\title{$O(1/N^3)$ conformal bootstrap solution of the $SU(2)$ $\times$ $SU(2)$
Nambu--Jona-Lasinio model.}
\author{J.A. Gracey, \\ Department of Applied Mathematics and Theoretical
Physics, \\ University of Liverpool, \\ P.O. Box 147, \\ Liverpool, \\
L69 3BX, \\ United Kingdom.\thanks{Address after 1st October, 1994:
Department of Mathematical Sciences, Science Laboratories, University of
Durham, South Road, Durham, DH1 3LE, United Kingdom.}}
\date{}
\maketitle
\vspace{5cm}
\noindent
{\bf Abstract.} Using the full conformal bootstrap method an analytic
expression is given in $d$-dimensions for the anomalous dimension of the
fermion at $O(1/N^3)$ in a large $N$ expansion of the Nambu--Jona-Lasinio model
with $SU(2)$ $\times$ $SU(2)$ continuous chiral symmetry.
\newpage
Four fermi interactions have become important recently in various areas of
field theory and its applications to particle and condensed matter physics. For
example, in the former area it had been proposed that the Higgs mechanism of
the standard model and the consequent mass generation could be reproduced,
\cite{1,2,3}, by a top quark condensate which had its origins in an interaction
involving four top fields. Originally such models were introduced in a
different context by Nambu and Jona-Lasinio, \cite{4}, as a low energy
effective theory describing hadronic physics, \cite{5}. Although these
considerations are four dimensional, four fermi theories also in fact play a
role in understanding lower dimensional phenomena. For example, one interest in
the three dimensional model centres on trying to ascertain the effect such an
interaction induces in models describing high $T_c$ superconductivity as it is
believed to figure in that mechanism, \cite{6}. In another direction, numerical
simulations have been performed on the lattice for models with small numbers of
flavours to observe the onset of non-perturbative effects and critical
exponents have also been measured, \cite{7}. Analytic calculations using
several orders of the large $N$ expansion are in fairly good agreement with
such results, \cite{8}. Further, the four fermi model, the Gross Neveu model,
\cite{A}, is in the same universality class as the infra-red fixed point of the
Yukawa model in all dimensions $d$, $2$ $<$ $d$ $<$ $4$, \cite{B,C,3}. This
equivalence is verified through $\epsilon$-expansion techniques and knowledge
of the perturbative structure which is also used to gain improved estimates of
critical exponents. Indeed there has been recent work in this direction,
[11-22], where the $\epsilon$-expansion of the $O(N)$ Gross Neveu model and the
related Nambu--Jona-Lasinio models with various continuous chiral symmetries
have been examined from the point of view of establishing equivalences with
various discrete models, \cite{20}. Clearly one important and fundamental
ingredient in such a programme is the provision of as much information on the
quantum theory as is calculationally possible. Various tools exist to achieve
this, one of which is the aforementioned explicit perturbation theory coupled
with $\epsilon$-expansion techniques. A second is the large $N$ analysis, where
$N$ is the number of fundamental fields of the theory, in which one directly
calculates $d$-dimensional expressions for the critical exponents order by
order in $1/N$. Indeed there has been intense activity in this area in the last
few years, [11-22], with the most successful being the application of
Vasil'ev's
self consistency technique, \cite{21,22}, to four fermi theories, \cite{8},
allowing the calculation of critical exponents both to $O(1/N^2)$ and
$O(1/N^3)$, [11,12,16-19]. Moreover the beauty of the latter technique is that
since it provides results in $d$-dimensions it contributes information to the
various problems mentioned earlier {\em simultaneously}.

In this letter we {\em complete} the application of the {\em full} conformal
bootstrap programme to four fermi type theories, \cite{14,15}, by deriving
the $O(1/N^3)$ expression for the anomalous dimension of the fermion in the
Nambu--Jona-Lasinio model with $SU(2)$ $\times$ $SU(2)$ continuous chiral
symmetry. Such a calculation, in arbitrary dimensions, is necessary for
proving the equivalence of that model with the Gell-Mann--L\'{e}vy
$\sigma$ model, \cite{23}, which also possesses an $SU(2)$ chiral symmetry.
The latter has been used as an effective theory to describe nucleons. The
lagrangian of the theory we consider is, \cite{4},
\begin{equation}
L ~=~ i \bar{\psi}^{iI} \partialslash \psi^{iI} + \sigma \bar{\psi}^{iI}
\psi^{iI} + i \pi^a \bar{\psi}^{iI} \lambda^a_{IJ}\gamma^5 \psi^{iJ}
- \frac{1}{2g^2}(\sigma^2 + \pi^{a2})
\end{equation}
where $\psi^{iI}$ is the fermion field with $1$ $\leq$ $i$ $\leq$ $N$, $1$
$\leq$ $I$ $\leq$ $M$, $1$ $\leq$ $a$ $\leq$ $(M^2-1)$, $\lambda^a_{IJ}$ are
the generators of, for the moment $SU(M)$, $g$ is the coupling constant and
$\sigma$ and $\pi^a$ are auxiliary bosonic fields. The $3$-point vertex form
is used here as it is more appropriate for applying the conformal bootstrap.
We note our conventions are $\mbox{Tr}(\lambda^a \lambda^b)$ $=$ $4T(R)
\delta^{ab}$, $\lambda^a \lambda^a$ $=$ $4C_2(R) I$ and $f^{acd} f^{bcd}$ $=$
$C_2(G) \delta^{ab}$ with $T(R)$ $=$ $\half$, $C_2(R)$ $=$ $(M^2-1)/2M$ and
$C_2(G)$ $=$ $M$ for the group $SU(M)$. Although we have noted the lagrangian
for the more general case $SU(M)$ $\times$ $SU(M)$ we will only consider $M$
$=$ $2$ in detail here. The reason for this is that the models with $M$ $=$ $2$
and $M$ $>$ $2$ have distinct properties which only became evident in recent
$O(1/N^2)$ calculations, \cite{24}. For example, the anomalous dimensions of
the $\sigma$ and $\pi^a$ fields are only equal in the abelian case, $U(1)$
$\times$ $U(1)$, and for $M$ $=$ $2$. For $M$ $>$ $2$, the equality is not
present. Mathematically this division can be traced to the totally symmetric
tensor $d_{abc}$ which is zero for $SU(2)$ but not for $SU(M)$, $M$ $>$ $2$.
Further, in the calculation of the $\beta$-function exponent at $O(1/N^2)$,
\cite{24}, one cannot solve the self consistency equation to deduce a simple
expression for $M$ $>$ $2$ as compared to $M$ $=$ $2$ where a closed analytic
solution emerges naturally. At $O(1/N)$ the self consistency formalism which
was used,  quite correctly reproduced results in agreement with \cite{11}.
Therefore we will concentrate here on the calculation of the fermion anomalous
dimension for the model $M$ $=$ $2$ and comment on the situation with $M$ $>$
$2$ later.

As a preliminary we give our notation and first recall that the conformal
propagators of the fields of (1), which are the starting point of the bootstrap
method, are, in the asymptotic region of coordinate space $x$ $\rightarrow$
$0$, \cite{16},
\begin{equation}
\psi(x) ~\sim~ \frac{A\xslash}{(x^2)^\alpha} ~~,~~
\sigma(x) ~\sim~ \frac{B}{(x^2)^\beta} ~~,~~
\pi(x) ~\sim~ \frac{C}{(x^2)^\gamma}
\end{equation}
Here $A$, $B$ and $C$ are the amplitudes of the fields and the critical
indices are defined, through dimensional analysis, to be
\begin{equation}
\alpha ~=~ \mu ~+~ \half \eta ~~,~~
\beta ~=~ 1 ~-~ \eta ~-~ 2\Delta_\sigma ~~,~~
\gamma ~=~ 1 ~-~ \eta ~-~ 2\Delta_\pi ~~,~~
\end{equation}
where $d$ $=$ $2\mu$, $\eta$ is the fermion anomalous dimension which we
calculate to $O(1/N^3)$ here and $\Delta_\sigma$ and $\Delta_\pi$ are the
$3$-vertex anomalous dimensions which satisfy, \cite{16},
\begin{equation}
2\alpha + \beta ~=~ 2\mu + 1 - 2\Delta_\sigma ~~,~~
2\alpha + \gamma ~=~ 2\mu + 1 - 2\Delta_\pi
\end{equation}
As already noted $\Delta_\sigma$ $=$ $\Delta_\pi$ to $O(1/N^2)$ and in
particular, with $\Delta$ $\equiv$ $\Delta_\sigma$ $=$ $\Delta_\pi$, and, for
example, $\eta$ $=$ $\sum_{i=1}^\infty \eta_i/N^i$, \cite{11,18,24},
\begin{eqnarray}
\eta_1 &=& - \, \frac{\Gamma(2\mu-1)}{\Gamma(\mu+1)\Gamma(\mu)\Gamma(1-\mu)
\Gamma(\mu-1)} \nonumber \\
\Delta_1 &=& - \, \frac{\mu\eta_1}{4(\mu-1)} \nonumber \\
\eta_2 &=& \eta^2_1 \left[ \frac{(\mu-2)\Psi}{2(\mu-1)} + \frac{1}{2\mu}
+ \frac{2}{(\mu-1)} - \frac{3\mu}{4(\mu-1)^2} \right] \nonumber \\
\Delta_2 &=& - \, \frac{\mu\eta^2_1}{16(\mu-1)^2}
\left[\frac{}{} 3\mu(\mu-1)\Theta + 2(\mu-2)\Psi \right. \nonumber \\
&&+~ \left. \frac{(5\mu-1)(2\mu^2-5\mu+4)}{(\mu-1)} \right]
\end{eqnarray}
where $\Psi(\mu)$ $=$ $\psi(2\mu-1)$ $-$ $\psi(1)$ $+$ $\psi(2-\mu)$ $-$
$\psi(\mu)$, $\Theta(\mu)$ $=$ $\psi^\prime(\mu)$ $-$ $\psi^\prime(1)$
and $\psi(x)$ is the logarithmic derivative of the $\Gamma$-function.

 From (2) the asymptotic scaling forms of the $2$-point function, which are
also
required, are
\begin{eqnarray}
\psi^{-1}(x) ~\sim~ \frac{r(\alpha-1)\xslash}{A(x^2)^{2\mu-\alpha+1}}
\nonumber \\
\sigma^{-1}(x) ~\sim~ \frac{p(\beta)}{B(x^2)^{2\mu-\beta}} \\
\pi^{-1}(x) ~\sim~ \frac{p(\gamma)}{B(x^2)^{2\mu-\gamma}} \nonumber
\end{eqnarray}
where $p(\alpha)$ $=$ $a(\alpha-\mu)/[\pi^{2\mu}a(\beta)]$, $q(\alpha)$ $=$
$\alpha p(\alpha)/(\mu-\alpha)$ and $a(\alpha)$ $=$
$\Gamma(\mu-\alpha)/\Gamma(\alpha)$. These results, (5), have been deduced
using the self consistency approach of \cite{21,9} where one solves the
skeleton Schwinger Dyson equations with dressed propagators but undressed
vertices. For the $O(N)$ Gross Neveu model these results have been reproduced
{\em exactly} using the {\em full} conformal bootstrap programme which will be
used here, [28,29,30,19,16]. The difference in this latter approach is that one
solves instead the $3$-point vertex function whose equivalent Dyson
representation is in terms of graphs with both dressed propagators and now
dressed vertices. The latter feature reduces substantially the number of
Feynman diagrams needed to be computed even at $O(1/N^3)$. The general
structure for the present case is illustrated in fig. 1 where the wavy line
represents either the $\sigma$ or $\pi^a$ fields. The original bootstrap
equations for a $\phi^3$ style theory have been derived in \cite{25,27} and
extended to the fermion case in \cite{14,16}. Rather than derive them
explicitly for (1) we simply state them as there are no major obstacles in
extending \cite{15} for (1).

We denote by $V_\sigma$ and $V_\pi$ the values of the respective vertex
functions and each will be a function of the exponents $\alpha$, $\beta$ and
$\gamma$ as well as various combinations of the amplitudes $z$ $=$
$f_\sigma A^2B$ and $y$ $=$ $f_\pi A^2C$. Here $f_\sigma$ and $f_\pi$ denote
the amplitudes of the respective Polyakov conformal triangle, \cite{26},
whose origin is as follows. From (4) the sum of the exponents of the lines
meeting at a vertex are $2\mu$ $+$ $1$ $-$ $\Delta$, where $\Delta$ $=$
$O(1/N)$. Therefore, recalling the dimensionality of the integration measure
associated with a vertex, the overall dimension of any vertex is $-$ $\Delta$
which is non-zero. Consequently one cannot apply directly the conformal
integration technique known as uniqueness, \cite{28}, or conformal
transformations on the integral representation of the graphs in the expansion
contained in fig. 1. To circumvent this difficulty in the conformal bootstrap
solution, one replaces each vertex by a Polyakov conformal triangle, [28-30],
which is illustrated in fig. 2. The exponents $\tilde{a}$ and $\tilde{b}$ of
the internal lines comprising this triangle are chosen in such a way that each
internal vertex is unique or conformal. This will therefore allow the use of
the aforementioned  conformal techniques to calculate each Feynman diagram. As
a consequence of representing each vertex by such a triangle, carrying out an
integration leads to the observation that the result is proportional to
$1/\Delta$ which is a reflection of the deviation from uniqueness.

A further set of variables upon which $V_\sigma$ and $V_\pi$ depend are the
infinitesimal regularizing parameters $\epsilon$, $\epsilon^\prime$ and
$\delta$. These are required in the formal derivation of the bootstrap
equations, \cite{26}, to avoid intermediate infinities and are introduced by
setting $\alpha$ $\rightarrow$ $\alpha$ $+$ $2\delta$, $\beta$ $\rightarrow$
$\beta$ $+$ $2\epsilon$ and $\gamma$ $\rightarrow$ $\gamma$ $+$
$2\epsilon^\prime$. (In the situation where the regulators are non-zero, it is
still possible to choose the internal propagators of a Polyakov conformal
triangle to maintain uniqueness at each vertex.) Therefore each of the
vertex functions have the formal dependence $V_{\sigma , \pi}$ $=$
$V_{\sigma , \pi}(z,y,\alpha,\beta,\gamma;\delta,\epsilon,\epsilon^\prime)$.

Equipped with these vertex functions and (2) and (6), we now write down the
formal conformal bootstrap equations for (1) which will be solved to obtain
$\eta_3$. First,
\begin{eqnarray}
V_\sigma(z,y,\alpha,\beta,\gamma;0,0,0) &=& 1 \nonumber \\
V_\pi(z,y,\alpha,\beta,\gamma;0,0,0) &=& 1
\end{eqnarray}
which reflect the fact that the sum of the graphs on the right side of fig. 1
is unity. In practice (7) is used to fix the normalization and give $z$ and $y$
at successive orders, though we note $z_1$ $=$ $y_1$. Secondly, for general
$M$,
\begin{eqnarray}
r(\alpha-1) &=& zt \, \left. \frac{\partial V_\sigma}{\partial \delta} \right|
+ 4yu C_2(R) \, \left. \frac{\partial V_\pi}{\partial \delta}\right| \\
p(\beta) &=& 2NMzt \, \left. \frac{\partial V_\sigma}{\partial \epsilon}
\right| \\
p(\gamma) &=& 8NT(R)ut \, \left. \frac{\partial V_\pi}{\partial\epsilon^\prime}
\right|
\end{eqnarray}
where $|$ denotes setting all regulators to zero and
\begin{eqnarray}
t &=& \frac{\pi^{4\mu}a^2(\alpha-1)a(\tilde{b})a(\beta)}
{\Gamma(\mu)(\alpha-1)^2 (\tilde{a}-1)^2 a(\beta-\tilde{b})} \nonumber \\
u &=& \frac{\pi^{4\mu}a^2(\alpha-1)a(\tilde{c})a(\gamma)}
{\Gamma(\mu)(\alpha-1)^2 (\tilde{a}-1)^2 a(\gamma-\tilde{c})}
\end{eqnarray}
where $\tilde{a}$, $\tilde{b}$ and $\tilde{c}$ are the exponents of the
internal lines of the respective conformal triangles. Eliminating $t$ and $u$
gives
\begin{eqnarray}
r(\alpha-1) &=& \frac{p(\beta)}{2NM} \left. \frac{\partial V_\sigma}
{\partial \delta} \right| \bigg/ \left. \frac{\partial V_\sigma}{\partial
\epsilon} \right| \nonumber \\
&+& \frac{C_2(R)p(\gamma)}{2NT(R)} \left. \frac{\partial V_\pi}
{\partial \delta} \right| \bigg/ \left. \frac{\partial V_\pi}{\partial
\epsilon^\prime} \right|
\end{eqnarray}
Thus knowledge of the values of the vertex functions to $O(1/N^2)$ means one
can deduce $\eta_3$ from (12) as it occurs at the same order in the left side
of (12). One simplification occurs in the calculation of the derivatives with
respect to the regulators. As noted earlier each conformal triangle yields a
pole in the deviation from uniqueness of the vertex it represents. Therefore an
$n$-vertex graph contributing to $V_\sigma$, for example, has the structure
$h(\Delta,\epsilon,\epsilon^\prime,\delta)/\Delta^{n-2}(\Delta-\delta)
(\delta-\epsilon)$ where $h$ is a non-singular function. Thus,
\begin{equation}
\left. \frac{\partial V_\sigma}{\partial \delta} \right| \bigg/ \left.
\frac{\partial V_\sigma}{\partial \epsilon} \right| ~=~
\left[ 1 + \Delta \left. \frac{\partial V_\sigma}{\partial \delta}
\right|_{\mbox{res}}\right]
\left[ 1 + \Delta \left. \frac{\partial V_\sigma}{\partial \epsilon}
\right|_{\mbox{res}}\right]^{-1}
\end{equation}
where $\mbox{res}$ denotes the contribution from differentiating the residue
function $h$ of the regularized vertex function. This is important since it
is not possible to evaluate exactly all the Feynman diagrams at $O(1/N^2)$
but only the difference defined as,
\begin{equation}
\tilde{\Delta}V ~ \equiv ~ \left. \frac{\partial V}{\partial \delta} \right|
{}~-~ \left. \frac{\partial V}{\partial \epsilon} \right|
\end{equation}

To complete the calculation the explicit values for $\tilde{\Delta}V_\sigma$
and $\tilde{\Delta}V_\pi$ are required. The values of the basic topologies
have been calculated in \cite{13} and it is a straightforward exercise to
include the effects of the $SU(M)$ $\times$ $SU(M)$ symmetry. For
completeness we record that the values for both $V_\sigma$ and $V_\pi$ are the
same for $M$ $=$ $2$ but differ for $M$ $>$ $2$. For the former case which we
are solving here, we record
\begin{eqnarray}
\tilde{\Delta} \Gamma_2 &=& \frac{\mu\eta_1}{2(\mu-1)^2} \\
\tilde{\Delta} \Gamma_3 &=& \frac{5\mu\eta_1}{4(\mu-1)^2} \left[ \frac{}{}
(\mu-1)(2\mu-1) - (2\mu^2-5\mu+4)\Psi \right. \nonumber \\
&&~~~~~~~~~~~~~~~- \left. \frac{5(2\mu^2-5\mu+4)}{2(\mu-1)} \right] \\
\tilde{\Delta} \Gamma_4 &=& \frac{\mu\eta_1}{8} \!\! \left[ 3\Theta\Xi
+ \frac{3\Xi}{(\mu-1)^2} + \frac{2\mu\Psi}{(\mu-1)^2}
+ \frac{(\mu-8)\Theta}{(\mu-1)} + \frac{2(2\mu-3)}{(\mu-1)^3} \right] \\
\tilde{\Delta} \Gamma_5 &=& \frac{\mu\eta_1}{8}
\!\! \left[ 3\Theta\Xi + \frac{3\Xi}{(\mu-1)^2}
+ \frac{2\mu\Psi}{(\mu-1)^2} - \frac{8\Theta}{(\mu-1)} \right. \nonumber \\
&&~~~~~~~- \left. \frac{(2\mu-5)(\mu-2)}{(\mu-1)^3} \right]
\end{eqnarray}
where $\Phi(\mu)$ $=$ $\psi^\prime(2\mu-1)$ $-$ $\psi^\prime(2-\mu)$ $-$
$\psi^\prime(\mu)$ $+$ $\psi^\prime(1)$ and $\Xi(\mu)$ $=$ $I(\mu)$ $+$
$2/3(\mu-1)$ and $I(\mu)$ is related to a $2$-loop integral which cannot be
given in a closed form in terms of $\psi$-functions, \cite{22}.

Consequently after a little algebra we obtain the arbitrary dimensional
expression
\begin{eqnarray}
\frac{\eta_3}{\eta^3_1} &=& \left[ \frac{(\mu-2)^2\Psi^2}{2(\mu-1)^2}
- \frac{(\mu-2)^2\Phi}{8(\mu-1)^2} - \frac{3\mu^2\Theta\Psi}{8(\mu-1)}
- \frac{3\mu^2\Xi}{16(\mu-1)} \! \left( \! \Theta + \frac{1}{(\mu-1)^2}
\right) \right. \nonumber \\
&&+~ \left. \frac{\mu\Theta}{4(\mu-1)} \left( \frac{1}{\mu}
- \frac{3\mu}{2(\mu-1)} - \frac{(5\mu-4)}{4(\mu-1)} - \frac{1}{2}
- \frac{\mu(\mu-16)}{8(\mu-1)} \right) \right. \nonumber \\
&&+~ \left. \left( \frac{3}{2\mu} - \frac{5\mu}{8} - \frac{3}{16}
+ \frac{33}{16(\mu-1)} - \frac{71}{16(\mu-1)^2} + \frac{13}{16(\mu-1)^3}
\right) \Psi \right. \nonumber \\
&&+~ \left. \frac{1}{2\mu^2} - \frac{3}{\mu} - \frac{19}{16} - \frac{5\mu}{8}
+ \frac{3}{2(\mu-1)} \right. \nonumber \\
&&+~ \left. \frac{19}{8(\mu-1)^2} - \frac{33}{8(\mu-1)^3}
+ \frac{19}{16(\mu-1)^4} \right]
\end{eqnarray}
in addition to reproducing the results of (5). Explicitly in three dimensions
we deduce
\begin{equation}
\eta_3 ~=~ \frac{32}{27\pi^6} \left[ \frac{189}{2}\zeta(3) - 9\pi^2 \ln 2
- \frac{51}{4}\pi^2 + \frac{1157}{9} \right]
\end{equation}
where $\zeta(z)$ is the Riemann zeta function.

To conclude, we have now completed the full conformal bootstrap analysis for
several four fermi theories as far as is calculationally possible, ie
$O(1/N^3)$. It is worth noting for completeness, however, the situation with
other cases. In the model with continuous $U(1)$ $\times$ $U(1)$ chiral
symmetry, using the full conformal bootstrap approach discussed here, an
expression for $\eta_3$ cannot be obtained. In this instance the bootstrap
equations themselves become singular, which can be seen in several ways, but
stems from the fact that this model contains more symmetries than the case with
$M$ $=$ $2$ since one has additionally $\Delta_1$ $=$ $0$. Indeed it is this
vanishing of $\Delta$ at leading order which means that one cannot obtain
non-contradictory solutions for $z_1$ and $y_1$ from the vertex normalization
equations, (7), \cite{13}. As these are important variables for pushing the
bootstrap analysis to $O(1/N^3)$, it appears that one does not initially have a
consistent set of equations to solve. Alternatively one can see this phenomenon
by expressing the same equations in the general non-abelian case in terms of
the group Casimirs and then examining the limit to the abelian model. For the
non-abelian model with $M$ $>$ $2$ we have endeavoured to repeat the above
analysis. Like the abelian case it is not possible to deduce a value for
$\eta_3$. One complication with the analysis, which prevents the emergence of
an expression in arbitrary dimensions, is again related to the presence of the
non-zero tensor $d_{abc}$, similar to the problem which arose in the
calculation of the critical $\beta$-function slope, \cite{24}.

\vspace{0.4cm}
\noindent
{\bf Acknowledgement.} Some of the tedious algebra was performed using
{\rm REDUCE} version 3.4.

\newpage
\noindent
{\Large {\bf Figure Captions.}}
\begin{description}
\item[Fig. 1] Expansion of $3$-vertex.
\item[Fig. 2] Polyakov conformal triangle.
\end{description}
\end{document}